\documentclass[10pt,conference]{IEEEtran}
\IEEEoverridecommandlockouts\IEEEpubid{\makebox[\columnwidth]{ 978-1-6654-3540-6/22~\copyright~2022 IEEE \hfill} \hspace{\columnsep}\makebox[\columnwidth]{ }}

\usepackage{graphicx}
\usepackage[cmex10]{amsmath}
\usepackage{cases}
\usepackage[tight,footnotesize]{subfigure}
\usepackage{amsthm} 
\usepackage{cite}
\usepackage{amssymb}
\usepackage{algorithm}
\usepackage{algorithmic}
\usepackage{multirow}
\usepackage{amsmath}
\usepackage{xcolor}
\usepackage{CJK}
\usepackage{subeqnarray}
\usepackage{cases}
\usepackage{enumerate}
\usepackage{cuted}
\usepackage{booktabs}

\ifCLASSINFOpdf
\else
\fi

\begin{document}
	
	\title{STAR-RIS Aided Covert Communications}
	
	\author{Han Xiao$^*$, Xiaoyan Hu$^*$,
		Pengcheng Mu$^*$, Wenjie Wang$^*$, Tong-Xing Zheng$^*$, Kai-Kit~Wong$^\dagger$, Kun~Yang$^\ddagger$.\\
		$*$School of Information and Communication Engineering, Xi'an Jiaotong University, Xi'an 710049, China\\
		$\dagger$ Department of Electronic and Electrical Engineering, University College London, London WC1E7JE, U.K\\
		$\ddagger$ School of Computer Science and Electronic Engineering, University of Essex, Colchester CO43SQ, U.K\\
	}

	\maketitle
	
	\begin{abstract}
		This paper investigates the multi-antenna covert communications assisted by a simultaneously transmitting and reflecting reconfigurable intelligent surface (STAR-RIS). In particular, to shelter the existence of communications between transmitter and receiver from a warden, a friendly full-duplex receiver with two antennas is leveraged to make contributions to confuse the warden. Considering the worst case, the closed-form expression of the  minimum detection error probability (DEP) at the warden is derived and utilized as a covert constraint. Then, we formulate an optimization problem maximizing the covert rate of the system under the covertness constraint and quality of service (QoS) constraint with communication outage analysis. To jointly design the active and passive beamforming of the transmitter and STAR-RIS, an iterative algorithm based on globally convergent version of method of moving asymptotes (GCMMA) is proposed to effectively solve the non-convex optimization problem.
		Simulation results show that the proposed STAR-RIS-assisted scheme highly outperforms the case with conventional RIS.
	\end{abstract}
	
	\maketitle
	
	\vspace{-2mm}
	\section{Introduction}\label{sec:S1}
	Recently, the technology of covert communications has emerged as a new security paradigm and attracted significant research interests in both civilian and military applications\cite{yan19}, which can shelter the existence of communications between transceivers  and provide a higher level of security for wireless communication systems than physical layer security.
	As a breakthrough work, \cite{bash13} first proved that only $O(\sqrt n)$ bits information can be transmitted covertly and reliably from transmitters to receivers over $n$ channel uses.
	Actually, this conclusion is pessimistic since the intrinsic uncertainty of wireless channels and the background noise are not taken into account in the considered communication systems of \cite{bash13}.
	For example, \cite{goeckel15} and \cite{wang18} indicate that $O(n)$ bits information can be transmitted to the receiver when  eavesdroppers don't exactly know the background noise power or the channel state information (CSI).
	Besides, existing works also resort to other uncertainties, e.g., power-varying artificial noise \cite{Hu19}, uninformed jammer \cite{zheng21},    to enhance the performance of covert communications. 
	
	The aforementioned works validate the effectiveness of the covert communication techniques from different perspectives, however, they only investigate the single-antenna covert communication scenarios. 
	In fact, multi-antenna technologies are beneficial in improving the capacity and reliability of traditional wireless communications which are also conducive to enhancing the performance of covert communications. For example, \cite{chen21} investigates the potential covert performance gain brought by multi-antenna systems.
	Although multi-antenna technologies can enhance covertness of communications through improving the ability of transmissions and receptions, it cannot tackle the issues brought by the randomness of wireless propagation environment.
	
	To break through this limitation, reconfigurable intelligent surface (RIS) has recently emerged as a promising solution \cite{X.HU_TCOM21RIS, chen21enhancing,Wang21} and
	been leveraged in many wireless communication scenarios including covert communications \cite{chen21enhancing,Wang21}.
	It is worth noting that the RISs applied by the aforementioned works only reflect the incident signals which are limited to the scenarios that the transmitters and receivers locating at the same side of the RISs. 
	However, in practical cases, users may be on either side of RIS, and thus the flexibility and effectiveness of conventional RIS appear inadequate in these cases. To overcome this limitation, a novel technology called simultaneously transmitting and reflecting RIS (STAR-RIS) is further emerged \cite{han22artificial},
	which is capable of adjusting the reflected and transmitted signals by controlling the reflected and the transmitted coefficients simultaneously and help establish a more flexible full-space smart radio environment with $360^{\circ}$ coverage.
	Therefore, STAR-RIS possesses a huge application potential in wireless communications. 
	However, the investigation of leveraging STAR-RISs into wireless communication systems is still in its infancy stage.
	As for secure communication systems, only a small number of state-of-the-art works have utilized STAR-RISs to enhance the system secure performance \cite{han22artificial}.
	
	To our best knowledge, the application of STAR-RIS in covert communications has not been studied in existing works. This is the first work investigates a STAR-RIS assisted multi-antenna covert communication scenario. In this paper, we investigate the STAR-RIS assisted multi-antenna covert
	communication. Specifically, a STAR-RIS-assisted covert communication architecture is constructed through which the legitimate users located on both sides of the STAR-RIS can be simultaneously
	served. Based on the constructed system, the closed-form expressions of the minimum detection error probability (DEP) at the warden are analytically derived considering the worst-case scenario. Then, an optimization problem maximizing the covert rate under the covert communication constraint and the quality of service (QoS) constraint based on communication outage analysis is established by jointly optimizing the active and passive beamforming. To solve this optimization problem with the strongly coupled optimization variables, an iterative algorithm based on GCMMA is proposed. The effectiveness of  the proposed algorithm and STAR-RIS assisted covert scheme are validated by numerical results.
	\vspace{-4mm}\section{System Model}\label{sec:S2}
	In this paper, we consider a STAR-RIS-assisted covert communication system model, mainly consisting of a $M$-antenna BS transmitter (Alice) assisted by a STAR-RIS with $N$ elements, a covert user (Bob) and a  warden user (Willie) both equipped with a single antenna, and an assistant public user (Carol) with two antennas.
	It is assumed that Carol operates in the full-duplex mode where one antenna receives the transmitted singles from Alice and the other one transmits jamming signals to weaken Willie's detection ability.
	The STAR-RIS is deployed at the users' vicinity to enhance the end-to-end communications between Alice and the legal users Bob and Carol while confusing the detection of the warden user Willie.
	Without loss of generality, we consider a scenario that Bob and Carol locate on opposite sides of the STAR-RIS which can be served simultaneously by the reflected (T) and  transmitted (R) signals via STAR-RIS, respectively.
	
	The wireless communication channels from Alice to STAR-RIS, and from STAR-RIS to Bob, Carol, Willie are denoted as \textbf{H$_{\rm{AR}}$}=$\sqrt{\textit{l$_{\rm{AR}}$}}\textbf{G$_{\rm{AR}}$}$$\in\mathbb{C^{\textit{N$\times M$} }}$ and \textbf{h}$_{\rm{rb}}$=$\sqrt{\textit{l$_{\rm{rb}}$}}\textbf{g$_{\rm{rb}}$}$$\in\mathbb{C^{\textit{N$\times 1$}}}$, \textbf{h}$_{\rm{rc}}$=$\sqrt{\textit{l$_{\rm{rc}}$}}\textbf{g$_{\rm{rc}}$}$$\in\mathbb{C^{\textit{N$\times 1$} }}$, \textbf{h}$_{\rm{rw}}$=$\sqrt{\textit{l$_{\rm{rw}}$}}\textbf{g$_{\rm{rw}}$}$$\in\mathbb{C^{\textit{N$\times 1$} }}$, respectively.
	In particular, \textbf{G$_{\rm{AR}}$} and \textbf{g$_{\rm{rb}}$}, \textbf{g$_{\rm{rc}}$}, \textbf{g$_{\rm{rw}}$} are the small-scale Rayleigh fading coefficients. In addition, \textit{l$_{\rm{AR}}$} and \textit{l$_{\rm{rb}}$}, \textit{l$_{\rm{rc}}$}, \textit{l$_{\rm{rw}}$} are the large-scale path loss coefficients. As for the full-duplex  assistant user Carol, its self-interference channel can be modeled as $h_{\mathrm{cc}}=\sqrt{\phi}g_{\mathrm{cc}}$, where $g_{\mathrm{cc}}\sim \mathcal{C N}(0, 1)$,  $\phi \in[0,1]$ is the self-interference cancellation (SIC) coefficient determined by the performing efficiency of the SIC \cite{Hu19}.
	
	In this paper, we assume that the instantaneous CSI between STAR-RIS and Alice, Bob, Carol (i.e., $\mathbf{H}_{\mathrm{AR}}$, $\mathbf{h}_{\mathrm{rb}}$,  $\mathbf{h}_{\mathrm{rc}}$) is available at Alice, while only the statistical CSI between STAR-RIS and the Willie ($\mathbf{h}_{\mathrm{rw}}$) is known at Alice. In contrast, it is assumed that Willie is capable to know the instantaneous CSI of all the users, i.e., $\mathbf{h}_{\mathrm{rw}}$, $\mathbf{h}_{\mathrm{rb}}$ and $\mathbf{h}_{\mathrm{rc}}$, but can only access the statistical CSI of Alice, i.e., $\mathbf{H}_{\mathrm{AR}}$. The reasonability of assumptions made on CSI at Alice and Willie can be found in \cite{xiao2023simultaneously}.
	In addition, we assume that the power of the jamming signals, denoted as $P_\mathrm{j}$, follows the uniform distribution 
	with $P_{\mathrm{j}}^{\max }$ being the maximum power limit.
	It is assumed that Willie can only obtain the  of the jamming power, 
	and thus it is difficult for Willie to detect the existence of communications between Alice and Bob under the random jamming interference. 
	
	When Alice communicates with Bob and Carol with the help of STAR-RIS, the received signals at Bob and Carol can be respectively expressed as
	\vspace{-1.5mm}
	\begin{align}
		y_\mathrm{b}[k]=&\mathbf{h}_{\mathrm{rb}}^H \boldsymbol{\Theta}_\mathrm{r} \mathbf{H}_{\mathrm{AR}}\left(\mathbf{w}_\mathrm{b} s_\mathrm{b}[k]+\mathbf{w}_\mathrm{c}s_\mathrm{c}[k]\right)\notag\\
		&+\mathbf{h}_{\mathrm{rb}}^H \boldsymbol{\Theta}_\mathrm{t} \mathbf{h}_{\mathrm{rc}}^* \sqrt{P_\mathrm{j}} s_\mathrm{j}[k]+n_\mathrm{b}[k],\label{eq_rec_b}
			\end{align}
			\begin{align}
		y_\mathrm{c}[k]=&\mathbf{h}_{\mathrm{rc}}^H \boldsymbol{\Theta}_\mathrm{t} \mathbf{H}_{\mathrm{AR}}\left(\mathbf{w}_\mathrm{b} s_\mathrm{b}[k]+\mathbf{w}_\mathrm{c}s_\mathrm{c}[k]\right)\notag\\
		&+h_{\mathrm{cc}} \sqrt{P_\mathrm{j}} s_\mathrm{j}[k]+n_\mathrm{c}[k],\label{eq_rec_c}
	\end{align}
	where $k \in \mathcal{K}\triangleq\{1, \ldots, K\}$ denotes the index of each communication channel use with the maximum number of $K$ in a time \mbox{slot}.
	$\boldsymbol{\Theta}_\mathrm{r}=\operatorname{Diag}\Big\{\sqrt{\beta_\mathrm{r}^1} e^{\mathrm{j} \phi_\mathrm{r}^1}, \ldots, \sqrt{\beta_\mathrm{r}^N}e^{\mathrm{j}\phi_\mathrm{r}^N}\Big\}$ and
	$\boldsymbol{\Theta}_\mathrm{t}=\operatorname{Diag}\Big\{\sqrt{\beta_\mathrm{t}^1} e^{\mathrm{j} \phi_\mathrm{t}^1}, \ldots, \sqrt{\beta_\mathrm{t}^N}e^{\mathrm{j}\phi_\mathrm{t}^N}\Big\}$
	respectively indicate the STAR-RIS reflected and transmitted coefficient matrices, where $\beta_\mathrm{r}^n, \beta_\mathrm{t}^n\in[0,1]$, $\beta_\mathrm{r}^n+\beta_\mathrm{t}^n=1$ and $\phi_\mathrm{r}^n, \phi_\mathrm{t}^n\in[0,2\pi)$, for $\forall n \in \mathcal{N} \triangleq\{1,2, \ldots, N\}$.
	In addition, {$\mathbf{w}_\mathrm{b}\in\mathbb{C^{\textit{M$\times 1$} }}$ and $ \mathbf{w}_\mathrm{c}\in\mathbb{C^{\textit{M$\times 1$} }}$ are the precoding vectors at Alice for Bob and Carol, respectively.
	\section{Analysis on STAR-RIS-Assisted Covert Communications}\label{sec:S3}%
	\subsection{Covert Communication Detection Strategy at Willie}\label{sec:S3_P1}
	In this section, we detail the detection strategy of Willie for STAR-RIS-assisted covert communications from Alice to Bob. In particular, Willie attempts to judge whether there exists covert transmissions based on the received signal sequence $\{y_\mathrm{w}[k]\}_{k \in \mathcal{K}}$ in a time slot.
	Thus, Willie has to face a binary hypothesis for detection, which includes a null hypothesis, $\mathcal{H}_0$, representing that Alice only transmits public signals to Carol, and an alternative hypothesis, $\mathcal{H}_1$, indicating that Alice transmits both public signals and covert signals to Coral and Bob, respectively.
	Furthermore, the received signals at Willie based on the two hypotheses are given by
	\begin{align}
		\mathcal{H}_0: y_\mathrm{w}[k]=&\mathbf{h}_{\mathrm{rw}}^H \boldsymbol{\Theta}_\mathrm{r} \mathbf{H}_{\mathrm{AR}} \mathbf{w}_\mathrm{c}s_\mathrm{c}[k]+\mathbf{h}_{\mathrm{rb}}^H \boldsymbol{\Theta}_\mathrm{t} \mathbf{h}_{\mathrm{rc}}^* \sqrt{P_\mathrm{j}} s_\mathrm{j}[k]\notag\\&+n_\mathrm{w}[k], ~k\in \mathcal{K}, \label{eq_hypoH0_w}\\
		\mathcal{H}_1: y_\mathrm{w}[k]=&\mathbf{h}_{\mathrm{rw}}^H \boldsymbol{\Theta}_\mathrm{r} \mathbf{H}_{\mathrm{AR}} \mathbf{w}_\mathrm{b} s_\mathrm{b}[k]+\mathbf{h}_{\mathrm{rw}}^H \boldsymbol{\Theta}_\mathrm{r} \mathbf{H}_{\mathrm{AR}} \mathbf{w}_\mathrm{c} s_\mathrm{c}[k]\notag\\&+\mathbf{h}_{\mathrm{rb}}^H \boldsymbol{\Theta}_\mathrm{t} \mathbf{h}_{\mathrm{rc}}^* \sqrt{P_\mathrm{j}}s_\mathrm{j}[k]+n_\mathrm{w}[k], ~k\in \mathcal{K},\label{eq_hypoH1_w}
	\end{align}
	where $n_\mathrm{w}[k]\sim\mathcal{C N}(0, \sigma_\mathrm{w}^2)$ is the AWGN received at Willie. 
	We assume that Willie utilizes a radiometer to detect the  covert signals from Alice to Bob, owing to its properties of low
	complexity and ease of implementation.
	
	According to the working mechanism of the radiometer, the average power of the received signals at Willie in a time slot, i.e., $\overline{P}_\mathrm{w}=\frac{1}{K} \sum_{k=1}^K\left|y_\mathrm{w}[k]\right|^2$, is employed for statistical test. Similar to the existing works, (e.g., \cite{Wang21}), we assume that Willie uses infinite number of signal samples to implement binary detection, i.e., $K \rightarrow \infty$.
	Hence, the average received power at Willie $\overline{P}_\mathrm{w}$ can be asymptotically approximated as
	\begin{align}\label{eq_repower_w}
		\overline{P}_\mathrm{w}=\begin{cases}\left|\mathbf{h}_{\mathrm{rw}}^H \boldsymbol{\Theta}_\mathrm{r} \mathbf{H}_{\mathrm{AR}} \mathbf{w}_\mathrm{c}\right|^2+\left|\mathbf{h}_{\mathrm{rw}}^H \boldsymbol{\Theta}_\mathrm{t} \mathbf{h}_{\mathrm{rc}}^*\right|^2 P_\mathrm{j}+\sigma_\mathrm{w}^2, & \mathcal{H}_0, \\ \left\|\mathbf{h}_{\mathrm{rw}}^H \boldsymbol{\Theta}_\mathrm{r} \mathbf{H}_{\mathrm{AR}} \mathbf{w}\right\|^2+\left|\mathbf{h}_{\mathrm{rw}}^H \boldsymbol{\Theta}_\mathrm{t} \mathbf{h}_{\mathrm{rc}}^*\right|^2 P_\mathrm{j}+\sigma_\mathrm{w}^2, & \mathcal{H}_1,\end{cases}
	\end{align}
	where $\mathbf{w}=[\mathbf{w}_\mathrm{b}, \mathbf{w}_\mathrm{c}]$. Hence, Willie needs to analyze $\overline{P}_\mathrm{w}$ to decide whether the communication between Alice and Bob is under the  hypotheses of $\mathcal{H}_0$ or $ \mathcal{H}_1$, and its decision rule can be presented as $\overline{P}_\mathrm{w} \underset{\mathcal{D}_0}{\stackrel{\mathcal{D}_1}{\gtrless}} \tau_\mathrm{dt}$, where $\mathcal{D}_0$ (or $\mathcal{D}_1$) indicates the decision that Willie favors $\mathcal{H}_0$ (or $\mathcal{H}_1$), and $\tau_\mathrm{d t}>0$ is the corresponding detection threshold.
	
	In this paper, we adopt the DEP as Willie's detection performance metric and consider the worst case scenario that Willie can optimize its detection threshold to obtain the minimum DEP.
	\vspace{-2mm}\subsection{Analysis on Detection Error Probability}\label{sec:S3_P2}
	In this section, we first derive the analytical expressions for $P_{\mathrm{e}}$ in closed form, based on the distribution of $\overline{P}_\mathrm{w}$ under $\mathcal{H}_0$ and $\mathcal{H}_1$.
	In particular, the analytical expressions for $P_{\mathrm{e}}$ is derived as
	\begin{align}\label{eq_DEP_expression}
		P_\mathrm{e}=\begin{cases}
			1, & \tau_\mathrm{d t}<\sigma_\mathrm{w}^2, \\
			1+\psi+\frac{\lambda-\tilde{\lambda}}{\gamma P_\mathrm{j}^{\max}}, & \sigma_\mathrm{w}^2 \leq \tau_\mathrm{d t}<\sigma_\mathrm{w}^2+\gamma P_\mathrm{j}^{\max}, \\
			1+\psi+\frac{\lambda e^{\frac{\chi}{\lambda}}+\tilde{\lambda} e^{\frac{\chi}{\tilde{\lambda}}}}{\gamma P_\mathrm{j}^{\max}},&\tau_\mathrm{d t}\geq\gamma P_\mathrm{j}^{\max}+\sigma_\mathrm{w}^2 ,
		\end{cases}
	\end{align}
	where $\lambda=\left\|\mathbf{h}_{\mathrm{rw}}^H \boldsymbol{\Theta}_\mathrm{r}\right\|_2^2\mathbf{w}_\mathrm{c}^H \mathbf{w}_\mathrm{c}$,
	$\tilde{\lambda}=\left\|\mathbf{h}_{\mathrm{rw}}^H \boldsymbol{\Theta}_\mathrm{r}\right\|_2^2\big(\mathbf{w}_\mathrm{b}^H\mathbf{w}_\mathrm{b}+ $ $\mathbf{w}_\mathrm{c}^H \mathbf{w}_\mathrm{c}\big)$, and $\gamma=\left|\mathbf{h}_{\mathrm{rw}}^H \boldsymbol{\Theta}_\mathrm{t} \mathbf{h}_{\mathrm{rc}}^*\right|^2$, $\psi=\frac{\tilde{\lambda} e^{-\frac{\tau_\mathrm{d t}-\sigma_\mathrm{w}^2}{\tilde{\lambda}}}-\lambda e^{-\frac{\tau_\mathrm{d t}-\sigma_\mathrm{w}^2}{\lambda}}}{\gamma P_\mathrm{j}^{\max}}$, $\chi=-\tau_\mathrm{dt}+\sigma_\mathrm{w}^2+\gamma P_\mathrm{j}^{\max}$.
	Note that more derivation details plesase refer to \cite{xiao2023simultaneously}.
	
	It is important to point out that we consider the worst case scenario that Willie can optimize its detection threshold to minimize the DEP. By analyzing the expression of DEP, the closed-form solution of the optimal $\tau_\mathrm{dt}^*$ is given by
	\begin{equation}\label{eq_opt_DT}
		\tau_\mathrm{dt}^*=\frac{\tilde{\lambda} \lambda}{\tilde{\lambda}-\lambda} \ln \Delta+\sigma_\mathrm{w}^2 \in \left[\sigma_\mathrm{w}^2+\gamma P_\mathrm{j}^{\max},\infty\right),
	\end{equation}
	where $\Delta=\frac{e^{\frac{\gamma P_\mathrm{j}^{\max}}{\lambda}}-1}{e^{\frac{\gamma P_\mathrm{j}^{\max} }{\tilde{\lambda}}}-1}$ is a function of $\lambda$, $\tilde{\lambda}$ and $\gamma$.
	
	Substituting \eqref{eq_opt_DT} into \eqref{eq_DEP_expression} and adopting some algebraic manipulations, the analytical closed-form expression of the minimum DEP can be obtained as
	\begin{align}\label{eq_opt_DEP}
		P_\mathrm{e}^*=&
		1-\notag \\
		&\frac{\tilde{\lambda}\left(e^{\frac{\gamma P_\mathrm{j}^{\max}}{\tilde{\lambda}}}-1\right)(\Delta)^{\frac{\lambda}{\lambda-\tilde{\lambda}}}-\lambda\left(e^{\frac{\gamma P_\mathrm{j}^{\max}}{\lambda}}-1\right)(\Delta)^{\frac{\tilde{\lambda}}{\lambda-\tilde{\lambda}}}}{\gamma P_\mathrm{j}^{\max}}.	
	\end{align}
	
	\vspace{-2mm}Since Alice only knows the statistical CSI of channel $\mathbf{h}_\mathrm{rw}$, the average minimum DEP over $\mathbf{h}_\mathrm{rw}$, denoted as $\overline{P}_\mathrm{e}^*=\mathbb{E}_{\mathbf{h}_{\mathrm{rw}}}\left(P_\mathrm{e}^*\right)$, is usually utilized to evaluate the covert communications between Alice and Bob \cite{ chen21}.
	In \eqref{eq_opt_DEP}, $\lambda$, $\tilde{\lambda} $ and $\gamma $ are all random variables including $\mathbf{h}_\mathrm{rw}$, and thus they are coupled to each other, which makes it challenging to calculate the $\overline{P}_\mathrm{e}^*$ directly.
	To solve this problem, large system analytic techniques in \cite{evans00} and \cite{xiao2023_star} are utilized to remove the couplings among $\lambda$, $\tilde{\lambda}$ and $\gamma$, then we can obtain the asymptotic analytic result of $P_\mathrm{e}^*$.
	In particular, we first apply the large system analytic technique on $\lambda$,  then the asymptotic equality about $\lambda$ can be given as 
	\begin{equation}\label{eq_asy_lambda_1}
		\begin{aligned}[b]
			\lim _{N \rightarrow \infty} \frac{\left\|\mathbf{h}_{\mathrm{rw}}^H \boldsymbol{\Theta}_\mathrm{r}\right\|_2^2\varpi_\mathrm{c}}{N}  		&\stackrel{(a)}{\rightarrow} \frac{l_{\mathrm{rw}} \varpi_\mathrm{c}}{N} \operatorname{tr}\left(\boldsymbol{\Theta}_\mathrm{r} \boldsymbol{\Theta}_\mathrm{r}^H\right) =\frac{\lambda_\mathrm{a}}{N} ,
		\end{aligned}
	\end{equation}
	where the convergence (a) is due to \cite[Corollary 1]{evans00}.
	Here, $\varpi_\mathrm{c}=\mathbf{w}_\mathrm{c}^H\mathbf{w}_\mathrm{c}$, $\theta_\mathrm{r}=\operatorname{diag}(\boldsymbol{\Theta}_\mathrm{r})^H\operatorname{diag}(\boldsymbol{\Theta}_\mathrm{r})$,  and $\lambda_\mathrm{a}=l_{\mathrm{rw}}\varpi_\mathrm{c} \theta_\mathrm{r}$ is the asymptotic result of $\lambda$. Similarly, the asymptotic result  of $\tilde{\lambda} $ can be expressed as $\tilde{\lambda}_\mathrm{a}=l_{\mathrm{rw}} \theta_\mathrm{r}(\varpi_\mathrm{b}+\varpi_\mathrm{c})$, where  $\varpi_\mathrm{b}=\mathbf{w}_\mathrm{b}^H\mathbf{w}_\mathrm{b}$.
	
	With the results of $\lambda_\mathrm{a}$ and $\tilde{\lambda}_\mathrm{a}$ based on the large system analytic technique, the uncertainty of $\lambda$ and $\tilde{\lambda}$ can be removed from the perspective of Alice. Substituting $\lambda_\mathrm{a}$ and $\tilde{\lambda}_\mathrm{a}$ into \eqref{eq_opt_DEP}, we obtain the asymptotic analytical result of the minimum DEP $P_\mathrm{e}^*$ with respect to (w.r.t.) the random variable $\gamma$, which is given by
	\begin{equation}\label{eq_asy_DEP}
		P_\mathrm{ea}^*=1-\frac{l_{\mathrm{rw}}\theta_\mathrm{r}\varpi_\mathrm{b}}{\gamma P_\mathrm{j}^{\max}}(\Delta(\gamma))^{\frac{-\varpi_\mathrm{c}}{\varpi_\mathrm{b}}}\left(e^{\frac{\gamma P_\mathrm{j}^{\max}}{l_{\mathrm{rw}}\theta_\mathrm{r}(\varpi_\mathrm{b}+\varpi_\mathrm{c})}}-1\right).	\
	\end{equation}
	
	It is easy to verify that $\gamma=\left|\mathbf{h}_{\mathrm{rw}}^H \boldsymbol{\Theta}_\mathrm{t} \mathbf{h}_{\mathrm{rc}}^*\right|^2 \sim\exp(\lambda_\mathrm{rw})$ where $\lambda_\mathrm{rw}=\left\|\boldsymbol{\Theta}_\mathrm{t}\mathbf{h}_{\mathrm{rc}}^*\right\|^2_2$.
	By averaging $P^*_\mathrm{ea}$ over $\gamma$, we can get the average asymptotic analytical result of the minimum DEP as
	\begin{equation}\label{eq_avera_asy_DEP}
		\begin{aligned}
			\overline{P}_\mathrm{e a}^*= \mathbb{E}_\gamma\left(P_\mathrm{e a}^*\right)
			=\int_0^{+\infty}P_\mathrm{e a}^*\left(\gamma\right)\frac{e^{\frac{-\gamma}{\lambda_{\mathrm{rw}}}}}{\lambda_{\mathrm{rw}}}d \gamma.
		\end{aligned}
	\end{equation}
	Due to the existence of $\Delta(\gamma)$ in $P_\mathrm{ea}^*$, the integral in \eqref{eq_avera_asy_DEP} for calculating $\overline{P}_\mathrm{ea}^*$ over the random variable $\gamma$  is non-integrable. Therefore, the exact analytical expression for $\overline{P}_\mathrm{ea}^*$ is mathematically intractable.
	In order to guarantee the covert constraint $\overline{P}_\mathrm{ea}^*\geq 1-\epsilon$  always holds,
	we further adopt a lower bound of $\overline{P}_\mathrm{ea}^*$ to evaluate the covertness of communications.
	Specifically, we use a lower bound $\hat\Delta(\gamma)\triangleq e^{ \gamma P_\mathrm{j}^{\max}\left(\frac{\tilde{\lambda}_\mathrm{a}-\lambda_\mathrm{a}}{\tilde{\lambda}_\mathrm{a} \lambda_\mathrm{a}}\right)}$ to replace $\Delta(\gamma)$, a lower bound of $\overline{P}_\mathrm{ea}^*$ can be obtained, which is given by
	\begin{equation}\label{eq_low_avera_asy_DEP}
		\begin{aligned}[b]
			\hspace{-2mm}\hat{P}_\mathrm{e a}^*	=1+\frac{l_{\mathrm{rw}} \theta_\mathrm{r} \varpi_\mathrm{b}\left(\ln \frac{l_{\mathrm{rw}} \theta_\mathrm{r}\left(\varpi_\mathrm{b}+\varpi_\mathrm{c}\right)}{l_{\mathrm{rw}} \theta_\mathrm{r}\left(\varpi_\mathrm{b}+\varpi_\mathrm{c}\right)+P_\mathrm{j}^{\max} \lambda_{\mathrm{rw}}}\right)}{P_\mathrm{j}^{\max} \lambda_{\mathrm{rw}}} <  \overline{P}_{\mathrm{ea}}^*.
		\end{aligned}
	\end{equation}
	
	Therefore, in the following sections, $\hat{P}_\mathrm{e a}^*\geq 1-\epsilon$ will be leveraged as a tighter covert constraint to jointly design the active and passive
	beamforming variables of the system.
	\vspace{0mm}	\subsection{Analysis on Communication Outage Probability}
	Note that, the randomness introduced by the jamming signal power $P_\mathrm{j}$ and the self-interference channel $h_{\mathrm{c}c}$ of  Carol is possible to result in communication outages between Alice and Bob/Carol. In order to guarantee the QoS of communications, the communication outage constraints should be taken into consideration. Hence, when the required transmission rate between Alice and Bob (or Carol) is selected as $R_\mathrm{b}$ (or $R_\mathrm{c}$), the closed-form expressions of the communication outage probabilities at Bob and Carol can be obtained, which is derived as
	\begin{align}
		\delta_\mathrm{AB} =&\begin{cases}
			0,  &\Upsilon>P_\mathrm{j}^{\max},\\	
			1-\frac{\Upsilon}{ P_\mathrm{j}^{\max}}, &0\leq\Upsilon\leq P_\mathrm{j}^{\max},\\
			1,&\Upsilon<0,
		\end{cases}\label{outage_exp_AB}\\
		\delta_\mathrm{AC}=&\begin{cases}e^{-\frac{\Gamma}{\phi P_\mathrm{j}^{\max}}}+\frac{\Gamma}{\phi P_\mathrm{j}^{\max}}\operatorname{Ei}\left(-\frac{\Gamma}{\phi P_\mathrm{j}^{\max}}\right)	,&\Gamma\geq 0,\\
			1, &\Gamma<0,	
		\end{cases}\label{outage_exp_AC}
	\end{align}
	where $\Upsilon=\frac{\left|\mathbf{h}_{\mathrm{rb}}^H \boldsymbol{\Theta}_\mathrm{r} \mathbf{H}_{\mathrm{AR}} \mathbf{w}_\mathrm{b}\right|^2}{\left(2^{R_\mathrm{b}}-1\right)\left|\mathbf{h}_{\mathrm{rb}}^H \boldsymbol{\Theta}_\mathrm{t} \mathbf{h}_{\mathrm{rc}}^*\right|^2}-\frac{\left(\left|\mathbf{h}_{\mathrm{rb}}^H \boldsymbol{\Theta}_\mathrm{r} \mathbf{H}_{\mathrm{AR}} \mathbf{w}_ \mathrm{c}\right|^2+\sigma_\mathrm{b}^2\right)}{\left|\mathbf{h}_{\mathrm{rb}}^H \boldsymbol{\Theta}_\mathrm{t} \mathbf{h}_{\mathrm{rc}}^*\right|^2}$, $\Gamma=\frac{\left|\mathbf{h}_{\mathrm{rc}}^H\boldsymbol{\Theta}_\mathrm{t} \mathbf{H}_{\mathrm{AR}} \mathbf{w}_ \mathrm{c}\right|^2}{\left(2^{R_\mathrm{c}}-1\right)}-\sigma_\mathrm{c}^2-\left|\mathbf{h}_{\mathrm{rc}}^H\boldsymbol{\Theta}_\mathrm{t} \mathbf{H}_{\mathrm{AR}} \mathbf{w}_ \mathrm{b}\right|^2$ and $\operatorname{Ei}(\cdot)$ is the exponential internal function given by $\operatorname{Ei}(x)=-\int_{-x}^{\infty}\frac{e^{-t}}{t} d t$.
	
	The communication outage constraints are then defined as $\delta_\mathrm{AB}\leq\iota$ and $\delta_\mathrm{AC}\leq\kappa$ where  $\iota$ and $\kappa$ are two communication outage thresholds required by the system performance indicators for Bob and Carol, respectively.
	In this paper, we try to maximize the covet rate of Bob under the covert constraint $\hat{\mathrm{P}}_\mathrm{e a}^*\geq 1-\epsilon$ and the communication outage constraints. In order to guarantee the two communication outage constraints, the upper bounds of $R_\mathrm{b}$ and $R_\mathrm{c}$, i.e., $R_\mathrm{bb}$  and $R_\mathrm{cc}$, are selected to represent the covert performance at Bob and QoS at Carol, respectively. The expressions of $R_\mathrm{bb}$  and $R_\mathrm{cc}$ is given by
	\begin{align}
		R_\mathrm{bb}=&\log _2 \left(1+
		\frac{\left|\mathbf{h}_{\mathrm{rb}}^H \boldsymbol{\Theta}_\mathrm{r} \mathbf{H}_{\mathrm{AR}} \mathbf{w}_ \mathrm{b}\right|^2}{\left|\mathbf{h}_{\mathrm{rb}}^H \boldsymbol{\Theta}_\mathrm{r} \mathbf{H}_{\mathrm{AR}} \mathbf{w}_ \mathrm{c}\right|^2+\widehat{\sigma}+\sigma_\mathrm{b}^2}\right),\label{eq_outrate_b}\\
		R_\mathrm{cc}=&\log _2 \left(1+\frac{\left|\mathbf{h}_{\mathrm{rc}}^H \boldsymbol{\Theta}_\mathrm{t} \mathbf{H}_{\mathrm{AR}} \mathbf{w}_\mathrm{c}\right|^2}{\left|\mathbf{h}_{\mathrm{rc}}^H \boldsymbol{\Theta}_\mathrm{t} \mathbf{H}_{\mathrm{AR}} \mathbf{w}_\mathrm{b}\right|^2+\sigma^*+\sigma_\mathrm{c}^2}\right), \label{eq_outrate_c}
	\end{align}
	where $\widehat{\sigma}=\left|\mathbf{h}_{\mathrm{rb}}^H \boldsymbol{\Theta}_\mathrm{t} \mathbf{h}_{\mathrm{rc}}^*\right|^2 P_\mathrm{j}^{\max}(1-\iota)$, $\sigma^*$  is the solution to the equation of $\delta_\mathrm{AC}=\kappa$ which can be numerically solved by the bi-section search method.
	\section{Problem Formulation and Algorithm Design}\label{sec:S4}
	\subsection{Optimization Problem Formulation and Reformulation}\label{sec:S4_P1}
	On the basis of the previous discussions in section \ref{sec:S3}, we formulate the optimization problem in this section. Specifically,
	we will maximize the covert rate between Alice and Bob under the covert communication constraint while ensuring the
	QoS at Carol with the QoS constraint, by jointly optimizing the active and passive beamforming variables, i.e., $\mathbf{w}_\mathrm{b}$, $\mathbf{w}_\mathrm{c}$, $\boldsymbol{\Theta}_\mathrm{r}$ and $\boldsymbol{\Theta}_\mathrm{t}$, Hence, the optimized problem formulation is expressed by
\vspace{-2mm}	\begin{subequations}\label{eq_for_opt}
		\begin{align}
			&\max _{\mathbf{w}_\mathrm{b}, \mathbf{w}_\mathrm{c}, \boldsymbol{\Theta}_\mathrm{r}, \boldsymbol{\Theta}_\mathrm{t} }R_\mathrm{bb},\notag\\
			&\qquad\text { s.t. }\left\|\mathbf{w}_\mathrm{b}\right\|_2^2+\left\|\mathbf{w}_\mathrm{c}\right\|_2^2 \leq P_{\text {tmax }}\label{eq_for_opt_1} \\
			&\qquad\qquad\hat{P}_\mathrm{e a}^*\geq 1-\epsilon ,\label{eq_for_opt_2}\\
			&\qquad\qquad R_\mathrm{cc}\geq R^*, \label{eq_for_opt_3}\\
			&\qquad\qquad\beta_\mathrm{r}^n+\beta_\mathrm{t}^n=1; \phi_\mathrm{r}^n, \phi_\mathrm{t}^n \in[0,2 \pi),\label{eq_for_opt_4}
		\end{align}
	\end{subequations}
	where \eqref{eq_for_opt_1} is the transmission power constraint for Alice with $P_\mathrm{tmax}$ being the maximum transmitted power; \eqref{eq_for_opt_2} is an equivalent covert communication constraint of $\hat{P}_\mathrm{e a}^*\geq 1-\epsilon$; 
	\eqref{eq_for_opt_3} represents the QoS constraint for Carol; 
	\eqref{eq_for_opt_4} is the amplitude and phase shift constraints for STAR-RIS.
	Actually, it is challenging to solve the formulated optimization problem because of the strong coupling among variables.
	To tackle this issue, an iterative algorithm, which includes an outer iteration and an inner iteration, based on GCMMA algorithm in \cite{svanberg02} is proposed to effectively solve the optimized problem \eqref{eq_for_opt}.
	It should be noted that the effectiveness of the GCMMA algorithm as presented in \cite{svanberg02} has been demonstrated through its ability to converge towards the Karush-Kuhn-Tucker (KKT) point of the original non-convex problem, provided that the KKT points is existed in the optimized problem.
	
	Specifically, we first equivalently reformulate the original optimization problem by introducing auxiliary variables $\mathbf{y}=[y_1, y_2, y_3]$ whose dimension is determined by the number of constraints in the original problem and $z$, and non-negative constants $a_0$ and $\mathbf{c}$. The transformed optimized problem is given by
		\vspace{-4mm}\begin{subequations}\label{eq_ori_trans}
		\begin{align}
			& \min _{\mathbf{V}, z, \mathbf{y}} f_0+a_0 z+\mathbf{c y},\notag \\
			& ~~\text { s.t. } f_i-y_i \leq 0, i\in\{1, 2, 3\} \label{eq_ori_trans_2}\\
			&\quad~~~~ y_i \geq 0, i=1,2,3 ; z \geq 0, \label{eq_ori_trans_3}
		\end{align}
	\end{subequations}
	where $\mathbf{V}=\left\{\mathbf{w}_\mathrm{b}, \mathbf{w}_\mathrm{c}, \boldsymbol{\Theta}_\mathrm{r}, \boldsymbol{\Theta}_\mathrm{t}\right\}$, $f_i, i\in {0, 1, 2, 3}$ is expressed as
\vspace{-2mm}	\begin{align}
		f_i(\mathbf{V})=\begin{cases}
			-R_\mathrm{bb}, & i=0,\\
			\left\|\mathbf{w}_\mathrm{b}\right\|^2+ \left\|\mathbf{w}_\mathrm{c}\right\|^2-P_\mathrm{tmax}, & i=1,\\
			1-\hat{P}_\mathrm{e a}^*-\epsilon, & i=2,\\
			-R_\mathrm{cc}+R^*, & i=3.
		\end{cases}
	\end{align}
\subsection{Proposed Iterative Algorithm}
	Next, we choose to solve the problem \eqref{eq_ori_trans}. Algorithm 1 concludes the basic iterative process of the proposed algorithm. In particular, we first apply the method of moving asymptotes (MMA) convex approximation in \cite{svanberg02} to transform the problem \eqref{eq_ori_trans} as a convex optimization problem, which is expressed as
	\vspace{-2mm}	\begin{subequations}\label{eq_ori_trans_approxi}
		\begin{align}
			& \min _{\mathbf{x}, z, \mathbf{y}} g_0^{(k, l)}(\mathbf{x})+a_0 z+\mathbf{c y},\notag \\
			& ~~\text { s.t. } g_i^{(k, l)}(\mathbf{x})-y_i \leq 0 , i\in\{1, 2, 3\} \label{eq_ori_trans_approxi_1}\\
			&\quad~~~~  \boldsymbol{\alpha}^{(k)}\leq\mathbf{x}\leq \boldsymbol{\beta}^{(k)},\label{eq_ori_trans_approxi_4}\\
			&\quad~~~~ \eqref{eq_ori_trans_3},\label{eq_ori_trans_approxi_5}
		\end{align}
	\end{subequations}
	where $\boldsymbol{\alpha}$ and $\boldsymbol{\beta}$ are the lower bound and upper bound for the variable $\mathbf{x}=\left[\boldsymbol{\omega}_\mathrm{b}^T,\boldsymbol{\omega}_\mathrm{c}^T, \boldsymbol{\varphi}_\mathrm{b}^T, \boldsymbol{\varphi}_\mathrm{c}^T, \boldsymbol{\beta}_\mathrm{r}^T, \boldsymbol{\varphi}_\mathrm{r}^T, \boldsymbol{\varphi}_\mathrm{t}^T \right]$. Note that $\mathbf{w}_\xi=\boldsymbol{\omega}_\xi\circ e^{j\boldsymbol{\varphi}_\xi}$, $\xi\in\{\mathrm{b}, \mathrm{c}\}$; $\boldsymbol{\vartheta}_\mathrm{r}=\operatorname{diag}( \boldsymbol{\Theta}_\mathrm{r})=\sqrt{\boldsymbol{\beta}_\mathrm{r}}\circ e^{j\boldsymbol{\varphi}_\mathrm{r}}$, $\boldsymbol{\vartheta}_\mathrm{t}=\operatorname{diag}( \boldsymbol{\Theta}_\mathrm{t})=\sqrt{1-\boldsymbol{\beta}_\mathrm{r}}\circ e^{j\boldsymbol{\varphi}_\mathrm{t}}$; $g^{(k, l)}_i(\mathbf{x})$, $i\in\{0, 1, 2, 3\}$ is the MMA convex approximation of objective or constraints in $k$-th outer loop iteration and $l$-th inner loop iteration and is given by
\vspace{-2mm}	\begin{align}
		g^{(k, l)}_i(\mathbf{x})=&\left(\mathbf{p}^{(k, l)}_i\right)^T\left(\mathbf{I} \oslash\left(\mathbf{U}^{(k)}-\mathbf{x}\right)\right)\notag\\&+\left(\mathbf{q}^{(k, l)}_i)^T(\mathbf{I} \oslash\left(\mathbf{x}-\mathbf{L}^{(k)}\right)\right)
		+r^{(k, l)}_i,
	\end{align}
	where $\mathbf{U}^{(k)}$ and $\mathbf{L}^{(k)}$ are the upper asymptote and lower asymptote; $\mathbf{p}^{(k. l)}_i $, $\mathbf{q}^{(k, l)}_i $ and $r^{(k, l)}_i$ are given by \cite{svanberg07}
	\begin{align}
		& \mathbf{p}^{(k, l)}_i=\operatorname{power}\left(\mathbf{U}^{(k)}-\mathbf{x}^{(k-1)}\right)\circ\Big(1.001\Big(\frac{\partial{f}_i}{\partial{\mathbf{x}}}\left(\mathbf{x}^{(k-1)}\right)\Big)^{+}\notag\\
		&+0.001\Big(\frac{\partial{f}_i}{\partial{\mathbf{x}}}\left(\mathbf{x}^{(k-1)}\right)\Big)^{-}+\rho^{(k, l)}_i\mathbf{I} \oslash\left(\mathbf{x}_{\max}-\mathbf{x}_{\min}\right)\Big),
	\end{align}
	\begin{align}
		& \mathbf{q}^{(k, l)}_i=\operatorname{power}\Big(\mathbf{x}^{(k-1)-}\mathbf{L}^{(k)}\Big)\circ
		\Big(0.001\Big(\frac{\partial{f}_i}{\partial{\mathbf{x}}}\left(\mathbf{x}^{(k-1)}\right)\Big)^{+}\notag\\&+1.001\Big(\frac{\partial{f}_i}{\partial{\mathbf{x}}}\left(\mathbf{x}^{(k-1)}\right)\Big)^{-}+\rho^{(k, l)}_i\mathbf{I} \oslash\left(\mathbf{x}_{\max}-\mathbf{x}_{\min}\right)\Big),
	\end{align}
	where operator $\circ$ and $\oslash$ respectively denote the element-wise product and element-wise division; $\operatorname{power}(\mathbf{a})$ represents a vector whose entries are the square of entries in $\mathbf{a}$; $\mathbf{I}$ is a vector whose all entries are 1; $\mathbf{x}_{\max}$ and $\mathbf{x}_{\min}$ are the range of $ \mathbf{x}$ in original problem; $\frac{\partial{f}_i}{\partial{\mathbf{x}}}$ is the gradient information of the objective function or constraints w.r.t. $\mathbf{x}$; $\rho_i^{(k, l)}$ denotes the conservative factor that controls the conservativeness of the approximated convex problem and when $l=0$, i.e., $\rho_i^{(k, 0)}$, the expression of the conservative factor is given by
	\begin{align}
		\rho_i^{(k, 0)}=\max\left\{\frac{0.1}{N_\mathrm{va}}\operatorname{abs}\left(\frac{\partial{f}_i}{\partial{\mathbf{x}}}\right)^T\left(\mathbf{x}_{\max}-\mathbf{x}_{\min}\right), 10^{-6}\right\},
	\end{align}
	where $\operatorname{abs}(\mathbf{a})$ denotes a vector whose entries are the absolute value of entries in $\mathbf{a}$.
\vspace{-4mm}
\begin{center}
	\begin{tabular}{p{8.5cm}}
		\toprule[2pt]
		\textbf{Algorithm 1:}  Proposed Iterative Algorithm Based on GCMMA  \\ 
		\midrule[1pt]
		1: Initialize $\mathbf{x}^{(0)}$ and $a_0$; Calculate $v$; Define the accuracy \\\quad~tolerance thresholds $\varepsilon$; Set iteration index $k=0$ for outer \\\quad~loop.\\
		2: \textbf{While} $v>\varepsilon$  or $k=0$ \textbf{do}                       \\
		3: \quad Set iteration index $l=0$ for inner loop iteration.\\
		4: \quad Calculate $\frac{\partial{f}_i}{\partial{\mathbf{x}}}\left(\mathbf{x}^{(k)}\right)$, $\mathbf{p}^{(k, l)}_i$, $\mathbf{q}^{(k, l)}_i$, $\mathbf{U}^{(k)}$, $\mathbf{L}^{(k)}$, $\rho_i^{(k, l)}$, \\\quad~~~$\boldsymbol{\alpha}^{(k)}$and $\boldsymbol{\beta}^{(k)}$ to establish the convex approximation \\\quad~~~optimization problem \eqref{eq_ori_trans_approxi}. \\
		5: \quad Solve problem \eqref{eq_ori_trans_approxi} to obtain $\mathbf{x}^{(l)}_\mathrm{t}$ and calculate $f_i(\mathbf{x}^{(l)}_\mathrm{t})$ \\\quad~~~and $g^{(k, l)}_i(\mathbf{x}^{(l)}_\mathrm{t})$. \\
		6:  \quad\textbf{While} $f_i(\mathbf{x}^{(l)}_\mathrm{t})>g^{(k, l)}_i(\mathbf{x}^{(l)}_\mathrm{t})$\textbf{do}  \\
		7:\qquad~~Update $\rho_i^{(k, l+1)}$with \eqref{rho_update} and construct more con-\\\qquad~~~~servative $g^{(k, l+1)}_i$.\\
		8: \quad~~~ Solve the approximated problem with updated \\\qquad~~~~$g^{(k, l+1)}_i$to acquire the $\mathbf{x}^{(l+1)}_\mathrm{t}$; Let $l=l+1$\\
		9: \quad\textbf{end while}\\
		10: \quad Update $\mathbf{x}^{(k+1)}$ with $\mathbf{x}^{(l)}_\mathrm{t}$; Calculate $v$ and let $k=k+1$ \\
		11: \textbf{end while}\\
		\bottomrule[2pt]
	\end{tabular}
\end{center}
	
	To ensure the maintenance of a monotonically decreasing or non-increasing objective value with the iterative index, a size evaluation between $f_i$ and $g^{(k, l)}_i$ is performed utilizing the solution $\mathbf{x}^{(l)}_\mathrm{t}$ derived from the approximated optimization problem to established a  more conservative approximated problem. Specifically, if $f_i(\mathbf{x}^{(l)}_\mathrm{t})>g^{(k, l)}_i(\mathbf{x}^{(l)}_\mathrm{t})$, the conservative factor  will be adjusted to construct a more conservative $g_i^{(k, l)}$, in other words, this updating process for $\rho_i^{(k, l)}$ aims to find a convex upper-bound approximation of the original problem. The updated expression for the conservative factor is given by
	\begin{align}\label{rho_update}
		\rho_i^{(k, l+1)}=\min\left\{1.1\left(\rho_i^{(k, l)}+\nu^{(k)}\right), 10\rho_i^{(k, l)}\right\},
	\end{align}
	where
	\begin{align}
		\nu^{(k)}=\frac{f_i\left(\mathbf{x}^{(l)}_\mathrm{t}\right)-g_i^{(k, l)}\left(\mathbf{x}^{(l)}_\mathrm{t}\right)}{h^{(k)}\left(\mathbf{x}^{(l)}_\mathrm{t}\right)}.
	\end{align}
	
	Here, $v>0$ represents the gap of objective values between two adjacent iterations, and the algorithm converges when $v$ is below a predefined accuracy threshold $\varepsilon$. For the detailed expression of $\mathbf{U}^{(k)}$, $\mathbf{L}^{(k)}$, $\boldsymbol{\alpha}^{(k)}$, $\boldsymbol{\beta}^{(k)}$ and $r_i^{(k, l)}$, please refer to \cite{svanberg02}.

	\vspace{-2mm}\subsection{Analysis on Computed Complexity}\label{sec:S4_P4}
	In this section, we give the analysis on the computational complexity of the proposed Algorithm 1. Specifically, the main complexity comes from calculating the partial derivative of the objective function and constraints w.r.t. variables and solving the convex approximated optimization problem by leveraging the primal-dual interior-point method (PDIPM)\footnote{To decrease the computed complexity, the operation of solving the inverse of the matrix isn't adopted to calculate the iterative direction in PDIPM. Instead, a method with lower calculated complexity is leveraged as presented in \cite{svanberg07}.}. According to the computational complexity of matrix and vector operations, the computed complexity of the outer iteration is dominated by $\mathcal{O}\left(NM+M^2\right)$. In terms of solving the approximation problem \eqref{eq_ori_trans_approxi}, the computational complexity can be computed as $\mathcal{O}\left(N+M\right)$, which is obtained by analyzing the operations of matrix and vector. Hence, the computational complexity of Algorithm 1 is $I_1\left(\mathcal{O}\left(NM+M^2\right)+(I_2+1)\mathcal{O}\left(N+M\right)\right)$, where $I_1$ denotes the total iteration number of the proposed algorithm, $I_2$ is the number of inner iterations.
	\section{Simulation Results}\label{sec:S5}
	In this section, we show the numerical simulation results to verify the effectiveness of the proposed STAR-RIS-assisted covert communication scheme implemented by the proposed optimization Algorithm 1. 
	Specifically, the large-scale path loss coefficient is modeled as $\frac{\rho_0}{d^{\alpha}}$, and we assume $\rho_0=-20$ dB, $\alpha=2.6$ and the distances are set as $d_\mathrm{AR}=50 m$, $d_\mathrm{rb}=20 m$, $d_\mathrm{rw}=15 m$ and $d_\mathrm{rc}=25 m$.
	Furthermore, we define the noise power $\sigma_\mathrm{b}^2=-100$ dBm, $\sigma_\mathrm{c}^2=-100$ dBm and the self-interference coefficient $\phi=-110$ dBm\cite{bharadia13}.
	To highlight the advantage of covert communication aided by STAR-RIS, we consider a baseline scheme which employs two adjacent conventional RISs to replace STAR-RIS where one is the reflection-only RIS and the other one is transmission-only RIS. We call this baseline scheme as "RIS-aided scheme".

	In Fig. \ref{fig:epsilonvsrate}, we investigate the influence of the covert requirement, i.e., $\epsilon$, on the performance of average covert rate, considering different $P_{\max}$. In particular, $P_{\max}=3$ dBw is selected to operate the RIS-aided baseline scheme for an evident comparison, and obvious performance improvement can be achieved by the proposed scheme. Even if a lower transmitted power budget, i.e., 0 dBw, is utilized, the proposed scheme can still obtain better performance. This is because the STAR-RIS possesses a more flexible regulation ability compared with the conventional RIS, which can adjust the element phases and amplitudes for both reflection and transmission.
	\begin{figure}[ht]
		\centering
		\includegraphics[scale=0.35]{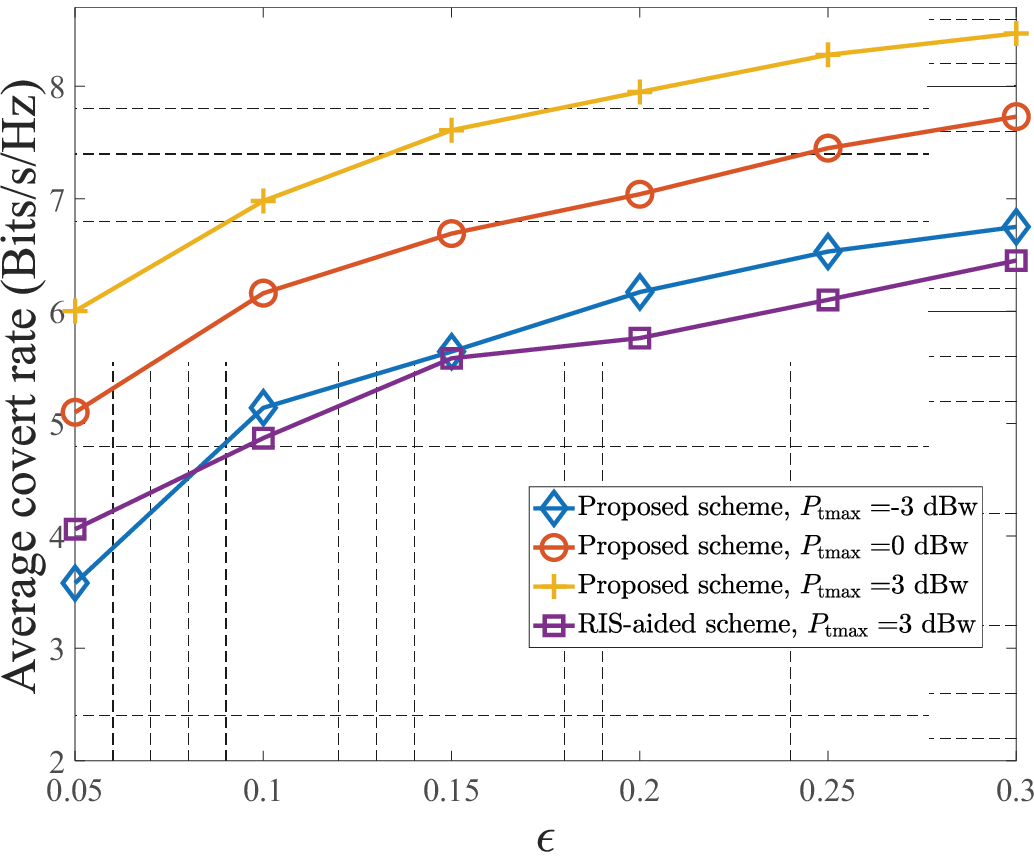}
		\caption{Average covert rate versus the covert requirement $\epsilon$ with $P_\mathrm{j}^{\max}=0$ dBw, $\iota=0.1$, $\kappa=0.1$, $M=3$, $N=30$, and $R^*=4$, and different $P_\mathrm{tmax}$.}\label{fig:epsilonvsrate}
	\end{figure}
	\begin{figure}[ht]
		\centering
		\includegraphics[scale=0.35]{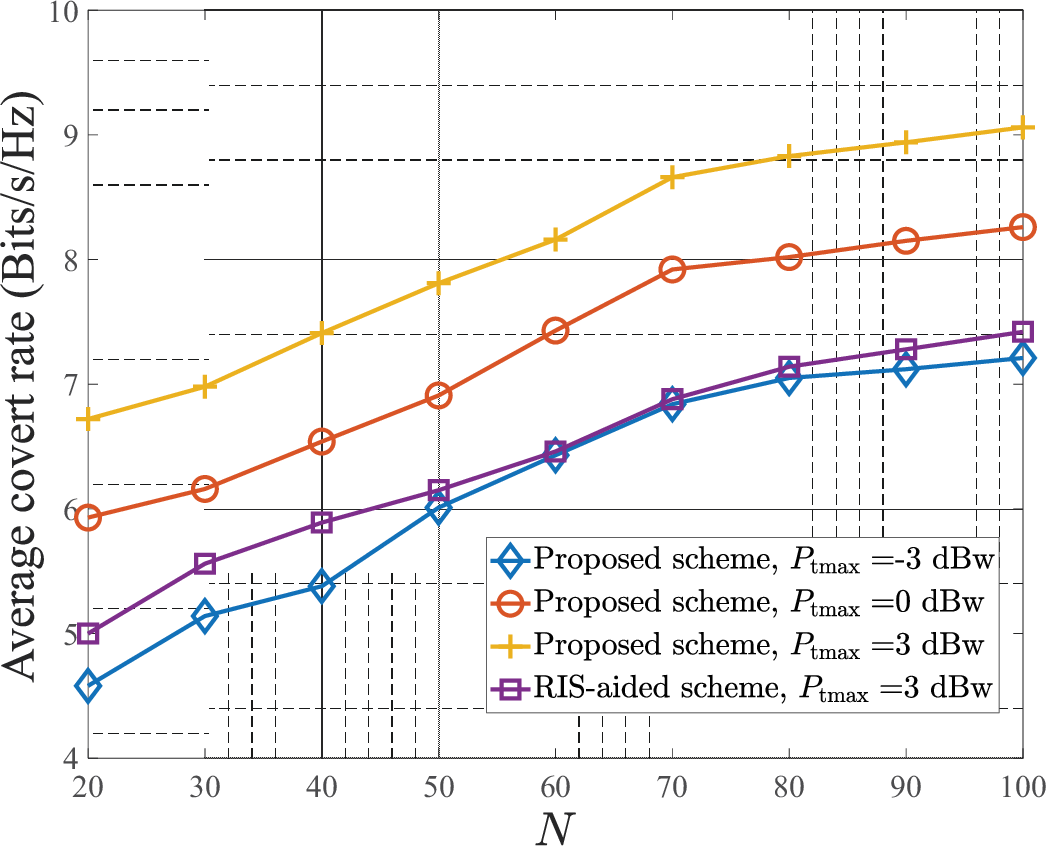}
		\caption{Average covert rate versus the number of antennas equipped at Alice with $P_\mathrm{j}^{\max}=0$ dBw, $\iota=0.1$, $\kappa=0.1$, $M=3$, and $R^*=4$, and different $P_\mathrm{tmax}$.}\label{fig:riselementvsrate}
	\end{figure}
	
	We present the variation curves of average covert rate w.r.t. the number of elements on STAR-RIS ($N$) in Fig. \ref{fig:riselementvsrate}, under different transmit power $P_\mathrm{tmax}$. It can be observed that the average covert rates of all the schemes grow with $N$, since the increased elements can provide higher freedom degree for reconfiguration of propagation environment. However, the increasing rates gradually decrease with the growth of $N$, and this may due to the limitations of other system settings. Similarly, $P_\mathrm{tmax}=3$ dBw is chosen to implement the two benchmark schemes, i.e., the RIS-aided scheme. 
	The obtained results further verify the advantages of the proposed STAR-RIS-assisted scheme which can achieve a performance that is not below the benchmark scheme in the scenario with a much smaller transmit power budget ($P_\mathrm{tmax}=-3$ dBw).

	%
	%
	\section{Conclusions}\label{sec:S6}
	In this work, we initially investigate the application potentials of STAR-RIS in covert communications. In particular, the closed-form expression of the minimum DEP about the STAR-RIS-aided covert communication system is analytically derived. And then we jointly design the active and passive beamforming at the BS and STAR-RIS to maximize the covert rate taking into account of the minimum DEP of Willie and the communication outage probability experienced at Bob and Carol. Due to the strong coupling between active and passive beamforming variables, the proposed optimization problem is a non-convex problem. To effectively solve this covert communication problem, we elaborately design an iterative algorithm based on GCMMA algorithm. Simulation results demonstrate that the STAR-RIS-assisted covert communication scheme highly outperforms the conventional RIS-aided scheme.
\vspace{-4mm}\section*{Acknowledgement}
		This work is partially supported by the National Natural Science Foundation of China under Grant 62201449, the Key R\&D Projects of Shaanixi Province under Grant 2023-YBGY-040, the Qin Chuang Yuan High-Level Innovation and Entrepreneurship Talent Program under Grant QCYRCXM-2022-231, the ``Si Yuan Scholar'' Foundation, the EU H2020 Project COSAFE under Grant GA-824019.

	\ifCLASSOPTIONcaptionsoff 
	\newpage
	\fi
	
	\bibliographystyle{IEEEtran}
	\bibliography{PSL-CC}

\begin{thebibliography}{10}
\providecommand{\url}[1]{#1}
\csname url@samestyle\endcsname
\providecommand{\newblock}{\relax}
\providecommand{\bibinfo}[2]{#2}
\providecommand{\BIBentrySTDinterwordspacing}{\spaceskip=0pt\relax}
\providecommand{\BIBentryALTinterwordstretchfactor}{4}
\providecommand{\BIBentryALTinterwordspacing}{\spaceskip=\fontdimen2\font plus
\BIBentryALTinterwordstretchfactor\fontdimen3\font minus
  \fontdimen4\font\relax}
\providecommand{\BIBforeignlanguage}[2]{{%
\expandafter\ifx\csname l@#1\endcsname\relax
\typeout{** WARNING: IEEEtran.bst: No hyphenation pattern has been}%
\typeout{** loaded for the language `#1'. Using the pattern for}%
\typeout{** the default language instead.}%
\else
\language=\csname l@#1\endcsname
\fi
#2}}
\providecommand{\BIBdecl}{\relax}
\BIBdecl

\bibitem{yan19}
S.~Yan, X.~Zhou, J.~Hu, and S.~V. Hanly, ``Low probability of detection
  communication: Opportunities and challenges,'' \emph{IEEE Wireless Commun.},
  vol.~26, no.~5, pp. 19--25, 2019.

\bibitem{bash13}
B.~A. Bash, D.~Goeckel, and D.~Towsley, ``Limits of reliable communication with
  low probability of detection on {AWGN} channels,'' \emph{IEEE J. Sel. Areas
  Commun.}, vol.~31, no.~9, pp. 1921--1930, 2013.

\bibitem{goeckel15}
D.~Goeckel, B.~Bash, S.~Guha, and D.~Towsley, ``Covert communications when the
  warden does not know the background noise power,'' \emph{IEEE Commun. Lett.},
  vol.~20, no.~2, pp. 236--239, 2015.

\bibitem{wang18}
J.~Wang, W.~Tang, Q.~Zhu, X.~Li, H.~Rao, and S.~Li, ``Covert communication with
  the help of relay and channel uncertainty,'' \emph{IEEE Wireless Commun.
  Lett.}, vol.~8, no.~1, pp. 317--320, 2018.

\bibitem{Hu19}
J.~Hu, S.~Yan, X.~Zhou, F.~Shu, and J.~Li, ``Covert wireless communications
  with channel inversion power control in rayleigh fading,'' \emph{IEEE Trans.
  Veh. Technol.}, vol.~68, no.~12, pp. 12\,135--12\,149, 2019.

\bibitem{zheng21}
T.-X. Zheng, Z.~Yang, C.~Wang, Z.~Li, J.~Yuan, and X.~Guan, ``Wireless covert
  communications aided by distributed cooperative jamming over slow fading
  channels,'' \emph{IEEE Trans. Wireless Commun.}, vol.~20, no.~11, pp.
  7026--7039, 2021.

\bibitem{chen21}
X.~Chen, W.~Sun, C.~Xing, N.~Zhao, Y.~Chen, F.~R. Yu, and A.~Nallanathan,
  ``Multi-antenna covert communication via full-duplex jamming against a warden
  with uncertain locations,'' \emph{IEEE Trans. Wireless Commun.}, vol.~20,
  no.~8, pp. 5467--5480, 2021.

\bibitem{X.HU_TCOM21RIS}
X.~Hu, C.~Masouros, and K.-K. Wong, ``Reconfigurable intelligent surface aided
  mobile edge computing: {F}rom optimization-based to location-only
  learning-based solutions,'' \emph{IEEE Trans. Commun.}, vol.~69, no.~6, pp.
  3709--3725, 2021.

\bibitem{chen21enhancing}
X.~Chen, T.-X. Zheng, L.~Dong, M.~Lin, and J.~Yuan, ``Enhancing {MIMO} covert
  communications via intelligent reflecting surface,'' \emph{IEEE Wireless
  Commun. Lett.}, vol.~11, no.~1, pp. 33--37, 2021.

\bibitem{Wang21}
C.~Wang, Z.~Li, J.~Shi, and D.~W.~K. Ng, ``Intelligent reflecting
  surface-assisted multi-antenna covert communications: Joint active and
  passive beamforming optimization,'' \emph{IEEE Trans. Commun.}, vol.~69,
  no.~6, pp. 3984--4000, 2021.

\bibitem{han22artificial}
Y.~Han, N.~Li, Y.~Liu, T.~Zhang, and X.~Tao, ``Artificial noise aided secure
  {NOMA} communications in {STAR-RIS} networks,'' \emph{IEEE Wireless Commun.
  Lett.}, 2022.

\bibitem{xiao2023simultaneously}
H.~Xiao, X.~Hu, P.~Mu, W.~Wang, T.-X. Zheng, K.-K. Wong, and K.~Yang,
  ``{Simultaneously Transmitting and Reflecting RIS (STAR-RIS) Assisted
  Multi-Antenna Covert Communications: Analysis and Optimization},''
  \emph{arXiv preprint arXiv:2305.04930}, 2023.

\bibitem{evans00}
J.~Evans and D.~N.~C. Tse, ``Large system performance of linear multiuser
  receivers in multipath fading channels,'' \emph{IEEE Trans. Inf. Theory.},
  vol.~46, no.~6, pp. 2059--2078, 2000.

\bibitem{xiao2023_star}
H.~Xiao, X.~Hu, A.~Li, W.~Wang, Z.~Su, K.-K. Wong, and K.~Yang, ``{STAR-RIS
  Enhanced Joint Physical Layer Security and Covert Communications for
  Multi-antenna mmWave Systems},'' \emph{arXiv preprint arXiv:2307.08043},
  2023.

\bibitem{svanberg02}
{K. Svanberg}, ``A class of globally convergent optimization methods based on
  conservative convex separable approximations,'' \emph{SIAM J. Optim.},
  vol.~12, no.~2, pp. 555--573, 2002.

\bibitem{svanberg07}
K.~Svanberg, ``{MMA} and {GCMMA}-two methods for nonlinear optimization,''
  \emph{Tech. Rep. Optim. Theory.}, vol.~1, 2007.

\bibitem{bharadia13}
D.~Bharadia, E.~McMilin, and S.~Katti, ``Full duplex radios,'' in \emph{Proc.
  ACM SIGCOMM}, 2013, pp. 375--386.

\end{thebibliography}

\end{document}